# Spices form the basis of food pairing in Indian cuisine

Anupam Jain[a,†], Rakhi N K[b,†] and Ganesh Bagler[b,*]

**Affiliations:**
[a]Centre for System Science, Indian Institute of Technology Jodhpur, Jodhpur, Rajasthan 342011, India.

[b]Centre for Biologically Inspired System Science, Indian Institute of Technology Jodhpur, Jodhpur, Rajasthan 342011, India.

[†]These authors contributed equally to this work
[*]Corresponding author: E-mail: bagler@iitj.ac.in, ganesh.bagler@gmail.com

## Abstract

Culinary practices are influenced by climate, culture, history and geography. Molecular composition of recipes in a cuisine reveals patterns in food preferences. Indian cuisine encompasses a number of diverse sub-cuisines separated by geographies, climates and cultures. Its culinary system has a long history of health-centric dietary practices focused on disease prevention and promotion of health. We study food pairing in recipes of Indian cuisine to show that, in contrast to positive food pairing reported in some Western cuisines, Indian cuisine has a strong signature of negative food pairing; more the extent of flavor sharing between any two ingredients, lesser their co-occurrence. This feature is independent of recipe size and is not explained by ingredient category-based recipe constitution alone. Ingredient frequency emerged as the dominant factor specifying the characteristic flavor sharing pattern of the cuisine. Spices, individually and as a category, form the basis of ingredient composition in Indian cuisine. We also present a culinary evolution model which reproduces ingredient use distribution as well as negative food pairing of the cuisine. Our study provides a basis for designing novel signature recipes, healthy recipe alterations and recipe recommender systems.

## Introduction

Culinary practices are shaped by complex interplay of culture, climate, geography and genetics (1–6). These factors influence food preferences and recipe composition, thereby altering the fabric of cuisine. Recipe composition pattern in a cuisine provides a means for investigating its gastronomic history and molecular constitution. The shift to cooked diet has been proposed to be a trigger for increased brain size in humans (7). Indian culinary system has traditionally developed dietary practices where food has nutritional as well as medicinal value (8–10). Ayurveda, the classic medicinal system of India, proposes that food has as much therapeutic value as drugs and even uses similar processing techniques for their preparation (11, 12).

In this study, we analyzed the recipe composition as well as flavor compound constitution of the Indian cuisine in search of its quintessential features. Specifically, we quantified the food pairing pattern and built models to identify features that explain statistical properties of the cuisine. The flavor constitution of the Indian cuisine was explored for ingredient composition and food pairing at the levels of cuisine, sub-cuisines, recipes and ingredient pairs. We built controls to probe for the role of factors that may be crucial in shaping recipes and hence the cuisine. Further we developed models to explain evolution of cuisine and its characteristic features. Our study illustrates application of data analysis and modeling for exploring chemical basis of a cuisine.

**Results**

**The Indian cuisine.** The traditional recipes of Indian cuisine are documented in the form of books (2) and lately through online repositories. We compiled the recipes data from TarlaDalal (13) (http://www.tarladalal.com, November 2014), one of the largest, most extensive online repertoires of original Indian recipes (*SI Text*). After curation, the Indian cuisine contained 2543 recipes from following eight sub-cuisines: Bengali, Gujarati, Jain, Maharashtrian, Mughlai, Punjabi, Rajasthani and South Indian. These sub-cuisines span diverse geographies, climates and cultures of Indian subcontinent. After aliasing ingredient names, the recipes comprised of 194 unique ingredients (*SI Text* and Dataset S1). The ingredients belong to following 15 categories: spice, vegetable, fruit, plant derivative, nut/seed, cereal/crop, dairy, plant, pulse, herb, meat, fish/seafood, beverage, animal product, and flower. Certain pulses (millets), lentils, spices and vegetables, seldom used in other cuisines, were common to Indian cuisine (*SI Text*).

Recipe size distribution reflects the ingredient richness of recipes in the cuisine. Similar to other cuisines (4, 14, 15), the recipe size distribution of Indian cuisine is bounded, varying between 1 and 40 with an average size of 7 (Fig. 1*A*). Mughlai sub-cuisine, with a royal lineage, had bimodal distribution with exceptionally large recipe sizes. Ingredient frequency when plotted against rank, by ordering ingredients according to their prevalence in the cuisine, reflects bias in use of ingredients. The frequency-rank distribution of the Indian cuisine varies over three orders of magnitude following a scale-free distribution (Fig. 1*B*). All sub-cuisines have strikingly similar profiles, indicative of generic culinary growth mechanisms (Fig 1*B*, inset). A few ingredients are excessively used, indicating either their inherent 'fitness' or possible accidental lock-in (15). Knowing that many cuisines share this property (4, 14, 15), the question is whether pattern of use of ingredients has any role in rendering the profile of a cuisine.

**Food pairing is measured in terms of overlap of flavor profiles.** The composition of recipes in a cuisine could be studied in terms of co-occurrence of ingredients (14, 16). One of the notions associated with ingredient co-occurrence is positive food pairing hypothesis— ingredients sharing flavor compounds are more likely to taste well together than ingredients that do not (14, 17). While this hypothesis holds true for some cuisines (North American, Western European and Latin American), it does not hold for a few others (Southern European and East Asian) that have negative food pairing (14). Thus, beyond following generic statistical patterns in recipe sizes as well as ingredient use, skewed food pairing seems to be a unique feature representing molecular basis of ingredient combinations dominant in a cuisine. Towards the aim of quantifying the pattern of ingredient composition of recipes, we studied food pairing (sharing of flavor compounds) in Indian cuisine.

While food sensation is a result of interplay between various aspects of ingredients, such as texture, color, temperature and sound, flavor plays a dominant role in specifying culinary fitness of ingredients and their combinations (18, 19). Flavor mediated food perception, primarily involving molecular interactions with olfactory and gustatory receptors, is crucial in developing food preferences in humans, and has coevolved with nutritional needs (20). Molecular composition of food dictates the sensation of flavor (21). Each ingredient is characterized by a set of chemical compounds which forms its flavor profile. Flavor profile provides us an effective tool for exploring patterns in ingredient composition of recipes. We obtained the flavor profiles for all ingredients in the Indian cuisine with the help of previously published data (14), a resource of flavor ingredients (21) and by extensive literature search (*SI Text*, Dataset S2). The flavor profiles, comprising of a total of 1170 unique compounds, had a size range of 270. Across this range of profile size, ingredient category representation was fairly uniform (Fig. S1).

The interrelationship among ingredients by virtue of shared flavor compounds could be represented as a flavor graph that illustrates the underlying topology of flavor sharing (Fig. S2 and S3). The ingredients have dominant intra-category flavor sharing indicating significant overlap of flavor profiles within the category (Fig. S4). We quantified flavor sharing in a recipe ($N_s^R$) and average flavor sharing of the cuisine ($\overline{N_s}$) by comparing profiles of ingredient pairs and their joint occurrence in recipes. Figure 2 illustrates this quantification procedure starting from data of recipes and flavor profiles

**Indian cuisine is characterized with strong negative food pairing.** We found that average flavor sharing in Indian cuisine was significantly lesser than expected by chance (Fig. 3*A*). When computed for all recipes in the cuisine, average flavor sharing for Indian cuisine was found to be 5.876, as compared to that of 9.442 for a randomized cuisine, which was constructed by randomly picking the ingredients while maintaining the recipe size distribution. This reflects a strong signature of non-random ingredient co-occurrence ($\Delta N_s = \overline{N}_s^{IC} - \overline{N}_s^{Rand}$ = -3.566 and Z-score of -54.727) skewed towards negative food pairing (Fig.3B). This is corroborated by the pattern of extent of flavor sharing between pairs of ingredients and their co-occurrence in the cuisine (Fig. 4*A*). More the extent of flavor sharing between any two ingredients in the Indian cuisine, lesser is their co-occurrence. The extent of food pairing bias in the Indian cuisine is much stronger than reported earlier for any other cuisine (14, 22) and is persistent regardless of the recipe size (Fig. 3*A*). Our analysis also showed that each of the sub-cuisines independently displayed negative food pairing, highlighting it as an invariant feature of the Indian cuisine (Fig. 3*B*). Thus, we conclude that the Indian cuisine is characterized with a strong negative food pairing.

We further explored the origin of this characteristic pattern by controlling for category and frequency of use of ingredients. The former is a recipe-level control that generates a cuisine by preserving category composition of each recipe, whereas the latter is a cuisine-level control that generates recipes by preserving frequency of occurrence of each ingredient. Interestingly, we observed that controlling only for the ingredient frequency leads to food pairing pattern similar to that of real-world cuisine (Fig. 3*A*, Fig. S5 and Fig. 4*B*). On the other hand, controlling only for ingredient category led to a pattern similar to that of a randomized cuisine. A randomized control that combines category-composition as well as ingredient frequency also reproduced the food pairing pattern. Thus, ingredient frequency emerged as the dominant factor specifying the characteristic flavor sharing pattern of the Indian cuisine. Considering the biased use of ingredients, we investigated the role of top ranked ingredients by randomly swapping the top ten ingredients with the rest. We found that indeed the highly ranked ingredients play a key role in shaping the negative food pairing pattern of the cuisine, in contrast to ingredients with poor ranking (Fig. S6).

**A copy-mutate model of the Indian cuisine explains the negative ingredient pairing.** Over the years, the present repertoire of recipes in Indian cuisine would have evolved from a much smaller primitive set of recipes. To probe for mechanisms and factors that may have influenced the cuisine, we implemented the copy-mutate model proposed by Kinouchi *et. al.*(15). This model imitates evolution of the cuisine to incorporate duplication and modification of recipes. The model with randomly ascribed ingredient fitness reproduced the frequency-rank distribution (Fig. 5*A*), but had food pairing similar to a random cuisine ($\overline{N_s}$ = 8.784, Fig. 5*C*). On the other hand, the cuisine generated using a modified copy-mutate model, in which the ingredient fitness was scaled to its occurrence, acquired the characteristic flavor sharing pattern of real-world cuisine, reproducing the negative food pairing ($\overline{N_s}$ = 6.183, Fig. 5 *C* and *D*). Interestingly, our model also suggests that the pattern of frequency-rank distribution may be a persistent feature throughout the evolution of the cuisine, with presence of a few dominant ingredients (Fig. 5*B*). The fitness parameter in our model could represent ingredient availability, flavor, nutritional value, cost and versatility (15).

**Spices are uniquely placed in the recipes.** Negative food pairing in the Indian cuisine is a cumulative result of individual ingredient contributions by virtue of pairing with other ingredients in recipes. To investigate the importance of individual ingredients and their categories in the composition of recipes, we randomized ingredients of each category independently, while maintaining the category as well as frequency of occurrence of the rest. We found that randomizing ingredients in any of the categories, except spices, does not affect negative food pairing pattern, thereby implying their insignificance (Fig. 6*A*). Spices, on the other hand, when swapped selectively, randomize the negative food pairing significantly (Fig. 6*A* and B, $\Delta N_s^{spice}$ = 4.229 and Z-score of -61.524). This implies that each of the spices is uniquely placed in its recipe to shape the flavor sharing pattern with rest of the ingredients, and is sensitive to replacement even with other spices, which is noteworthy given that the extent of flavor sharing is high among spices (Fig. S4 *A*).

**Spices are key contributors to the negative ingredient pairing.** Beyond global statistical features, we identified the ingredients that make key contributions towards the food pairing by computing the extent to which their presence affects the magnitude of average food pairing ($\chi$) . We found that the key ingredients that contribute to negative food pairing of Indian cuisine were spices (Fig. 6*C*). Among the top ten ingredients whose presence bias flavor sharing pattern of the Indian cuisine towards negative pairing, nine were spices: cayenne, green bell pepper, coriander, garam masala, tamarind, ginger garlic paste, ginger, clove, and cinnamon (*SI Text* and Dataset S3).

We surmise that this pivotal role of spices carries the evidence of historical practice of health-centric diet in Indian subcontinent.

**Discussion.** Our study reveals that spices occupy a unique position in the ingredient composition of Indian cuisine and play a major role in defining its characteristic profile. Spices, individually and as a category, emerged as the most critical contributors to the negative food pairing. Historically, they have been used as functional foods to serve multiple purposes such as coloring and flavoring agents, preservatives, and additives (23). Spices also find mention as medicines in Ayurveda as described in texts such as *Charaka Samhita* (11, 12, 24). One of the strongest rationales for the use of spices is the antimicrobial hypothesis— spices are primarily used due to their activity against food spoilage bacteria (3, 25). A few of the most potent antimicrobial spices (26) are commonly used in Indian cuisine. Spices also serve as antioxidant, anti-inflammatory, chemopreventive, antimutagenic, and detoxifying agents (24, 27). Our recent studies have shown the beneficial role of capsaicin, an active component in cayenne which was revealed to be the most critical ingredient in rendering the food pairing of the cuisine (28). The significance of spices in Indian cuisine is also highlighted by the fact that its recipes have many derived ingredients that are spice combinations (*SI Text*). Archeological evidences have suggested to the fact that lentils, millets and spices, especially turmeric and garlic were used as ingredients in ancient Indus civilizations (29, 30). We conclude that the evolution of cooking driven by medicinal beliefs would have left its signature on traditional Indian recipes.

Traditionally, recipes have been passed down the generations via oral renditions. Documentation of recipes in the form of cookbooks enable preservation of culinary practices (2). Our copy-mutate model of culinary evolution incorporates aspects central to diversification of the cuisine and closely reproduces its flavor sharing characteristics at the level of cuisine, recipes as well as ingredient pairs. While we have examined Indian cuisine on the basis of one of the most comprehensive resources, there is ample scope to enhance the analysis with enriched data. The flavor profiles of ingredients are limited by the availability of data. Also, our study does not account for the fact that certain flavor compounds may undergo changes in the process of cooking. The study of molecular basis of the cuisine has potential to be extended into nutritional genomics to explore the role of dietary chemicals in health (31, 32). Beyond revealing the characteristic food pairing of Indian cuisine, our study could potentially lead to methods for creating novel Indian signature recipes, healthy recipe alterations and recipe recommender systems.

**Methods**

**Flavor sharing.** We enumerated the flavor sharing (14) pattern among the ingredients that co-occur in a recipe, starting from the set of 2,543 ($N_R$) traditional Indian recipes comprising of 194 ($N_I$) ingredients (*SI Text*). We computed average number of shared compounds in each recipe $N_s^R$ and further calculated a representative average flavor sharing index $\overline{N_s}$ ($= \sum_R N_s^R / N_R$) of the cuisine. Figure 2 presents a graphic illustration of this procedure. For a recipe R with *n* ingredients $N_s^R$ is defined as,

$$N_s^R = \frac{2}{n(n-1)} \sum_{i,j \epsilon R, i \neq j} |F_i \cap F_j|$$

where $F_i$ represents the flavor profile of ingredient $i$ (a set of compounds).

Average flavor sharing in Indian cuisine was compared with a corresponding randomized cuisine to assess its statistical relevance by computing $\Delta N_s = \overline{N}_s^{IC} - \overline{N}_s^{Rand}$, where 'IC' and 'Rand' indicate Indian cuisine and corresponding random cuisine, respectively. Five types of randomized cuisines were created by maintaining the recipe size distribution of the original Indian cuisine: a randomized control where ingredients were chosen uniformly (20,000 recipes); a frequency-preserved control in which frequency of use of ingredients was preserved (20,000 recipes); a category-preserved control in which while the category composition of the recipe was preserved, ingredients were randomly chosen from each constituent category (8 sets of control cuisines, 20,344 recipes); a frequency-and-category-preserved control where the category composition was maintained and each ingredient was chosen with probability consistent with its frequency in Indian cuisine (8 sets of control cuisines, 20,344 recipes); a frequency-preserved randomized control where the top 10 ranked ingredients in the Indian cuisine were randomly swapped with low ranked (rank $\geq$ 11) ingredients (10 sets of control cuisines, 200,000 recipes). The statistical significance of $\Delta N_s$ was measured with Z-score, $Z = \sqrt{N_{Rand}} \frac{(\overline{N}_s^{IC} - \overline{N}_s^{Rand})}{\sigma_{Rand}}$. Here $N_{Rand}$ and $\sigma_{Rand}$ represent total number of recipes in the randomized cuisine and standard deviation, respectively.

**Copy-mutate model.** Copy-mutate model (15) imitates evolution of a cuisine to incorporate duplication and modification of recipes. It was started with an initial set of recipes ($R_0 = 20$) of size $K$ (= 7) each, by random selection of ingredients from a randomly populated ingredient pool. This initial set was evolved by copying a recipe randomly, mutating it $L$(= 2) times and adding this mutated recipe back to the set. For every mutation an ingredient was randomly chosen from the recipe and was replaced with another ingredient chosen randomly from the pool, only if the latter had higher fitness value than the former. New ingredients were introduced for maintaining the size ratio of ingredient pool and recipe pool ($M = 0.0762$, consistent with that of Indian cuisine). This process was repeated to obtain 61,032 recipes comprising 24 sets of cuisines generated through copy-mutate model. While one of the models was implemented with the strategy of Kinouchi *et. al.* (15) with randomly ascribed fitness values to ingredients ('fitness random'), a derivative of the same had fitness values scaled to ingredient frequency in the Indian cuisine ('fitness ranked').

**Ingredient uniqueness.** Uniqueness of an ingredient of a given category by virtue of flavor sharing pattern with other ingredients in the recipe was computed by replacing it with a randomly chosen ingredient from the same category. Deviation in the average flavor sharing of the randomized recipes (8 sets of control cuisines, 20,344 recipes) from that of the original cuisine was measured, for 10 major categories (depicted in Fig. S4).

$$\Delta N_s^{cat} = |\overline{N}_s^{IC} - \overline{N}_s^{cat}| \; \forall \; s \geq 2$$

Here, *cat* stands for ingredient category. This index enumerates contribution of ingredients of a given category towards the flavor sharing pattern of the cuisine. The statistical significance of $\Delta N_s^{cat}$ was measured with the Z-score.

**Ingredient contribution.** The contribution of each ingredient ($\chi_i$) to the flavor sharing pattern of the cuisine was quantified (14) in terms of the extent to which its presence biases the flavor pairing.

$$\chi_i = \left( \frac{1}{N_R} \sum_{R \ni i} \frac{2}{n(n-1)} \sum_{j \neq i (j,i \in R)} |F_i \cap F_j| \right) - \left( \frac{2f_i}{N_R \langle n \rangle} \frac{\sum_{j \in c} f_j |F_i \cap F_j|}{\sum_{j \in c} f_j} \right)$$

Here, $f_i$ is the frequency of occurrence of ingredient $i$.

All computations were performed on Dell Precision T5610 workstations (*Sushruta, Panini*) of the Complex Systems Laboratory, IIT Jodhpur.

**Acknowledgments:** G.B. acknowledges the seed grant support from Indian Institute of Technology Jodhpur (IITJ/SEED/2014/0003). A.J. and R.N.K. thank the Ministry of Human Resource Development, Government of India as well as Indian Institute of Technology Jodhpur for scholarship and Junior Research Fellowship, respectively. G.B. thanks M. S. Valiathan, M. K. Unnikrishnan, C. Suguna, Girish Arabale, Sutirth Dey, Hemachander Subramanian, L. S. Shashidhara and Prasanta Panigrahi for critical comments and suggestions.

**Figures**

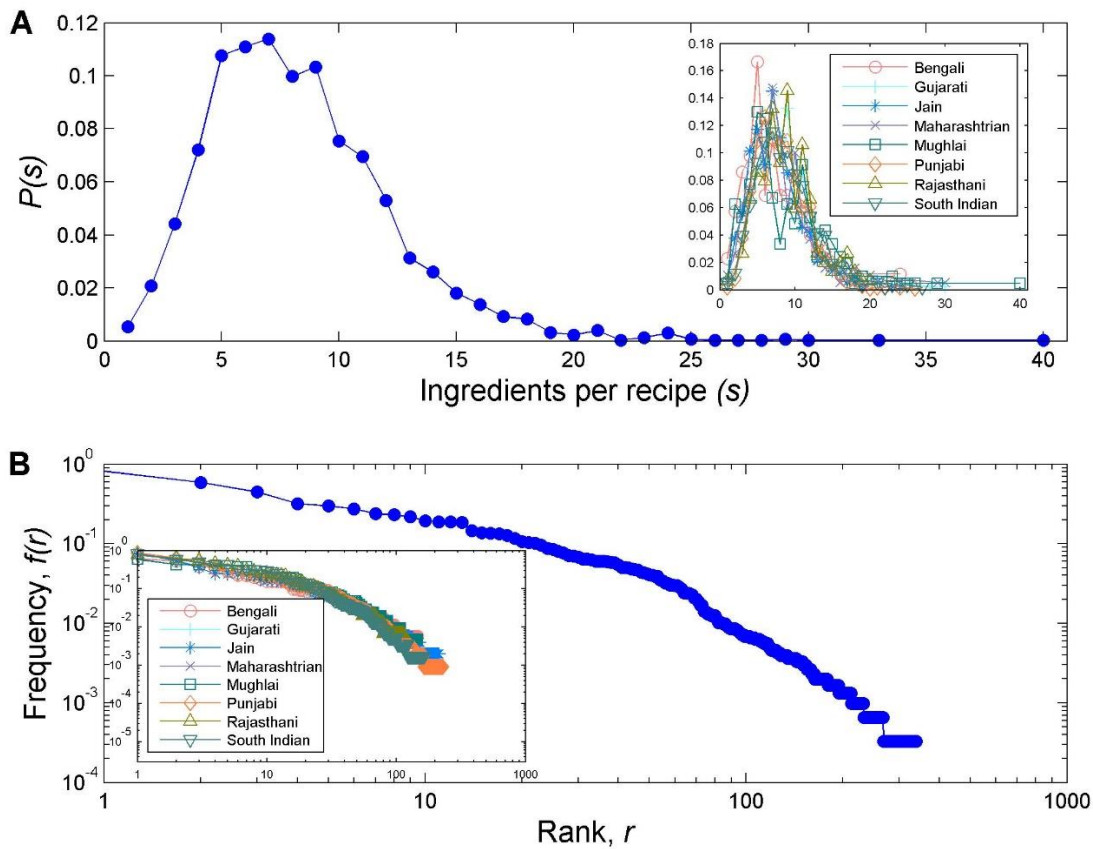

**Fig. 1. Statistics of Indian cuisine.** (**A**) Recipe size distribution of the Indian cuisine exhibits a bounded distribution similar to other cuisines (4, 14, 15). The inset shows distribution for constituent sub-cuisines. (**B**) Frequency rank plot of use of ingredients, in Indian cuisine and its sub-cuisines (inset), shows preferential use of few ingredients. The more frequently an ingredient is used in the cuisine the better is its rank.

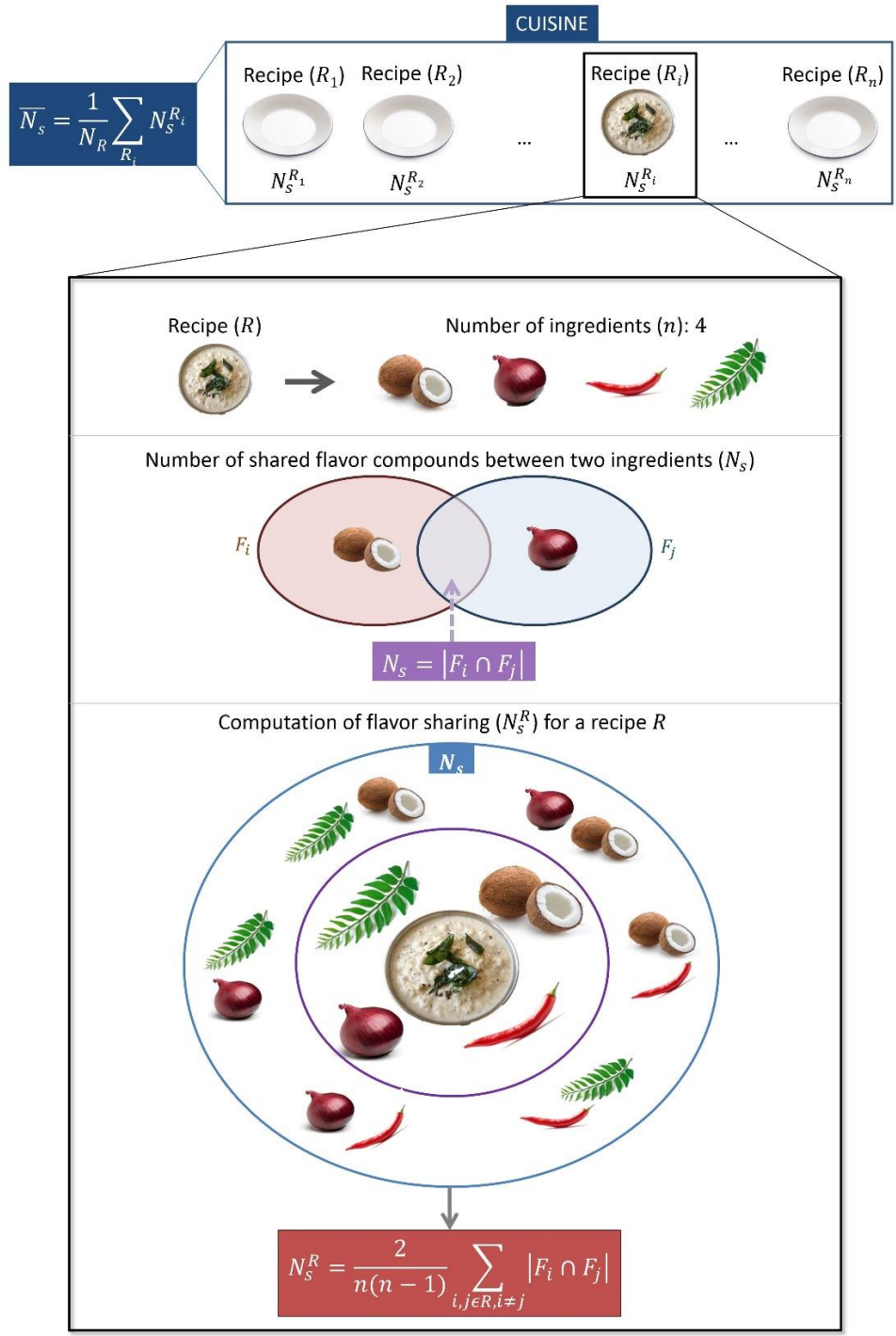

**Fig. 2. Illustration of procedure used for computation of average food pairing of a cuisine.** Starting from the cuisine data and flavor profiles of ingredients, average number of shared compounds in each recipe was computed. The average food pairing of a recipe set was further computed to enumerate flavor sharing.

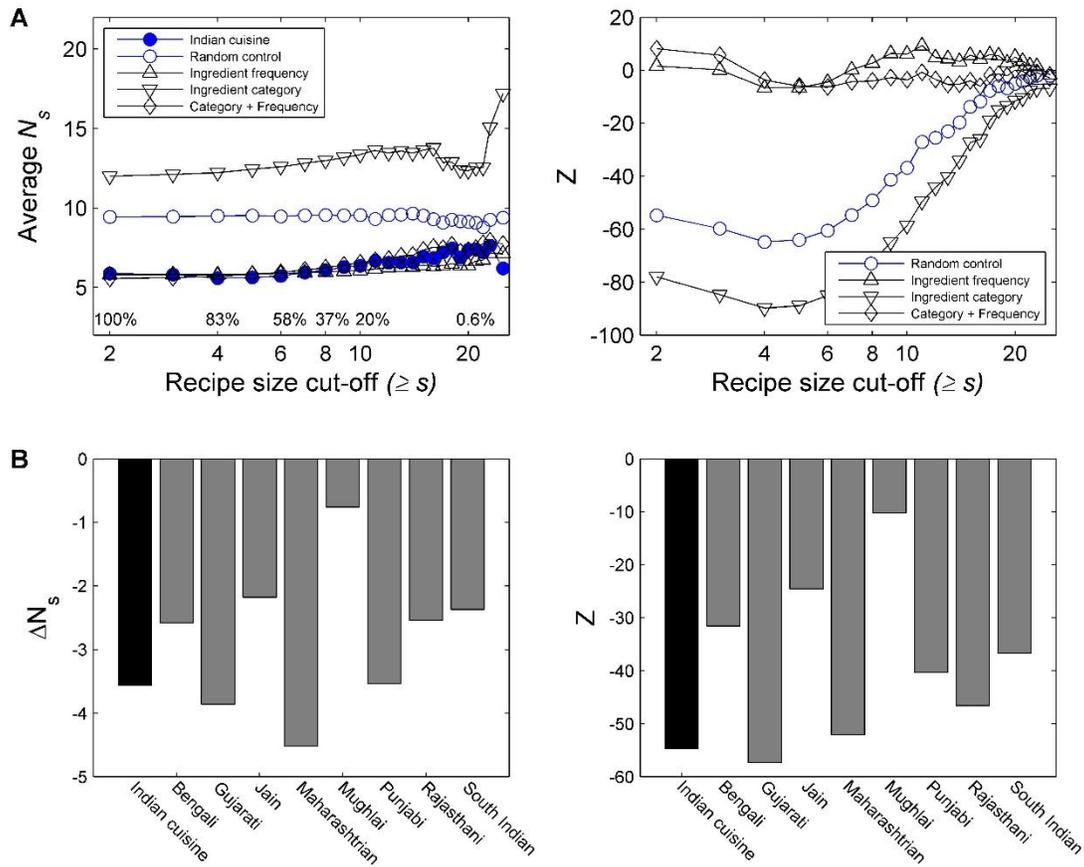

**Fig. 3. Strong negative food pairing in Indian cuisine and constituent sub-cuisines.** (**A**) The Indian cuisine is characterized by strong negative flavor sharing, when compared to its random control. The pattern of negative food pairing is independent of the recipe size ($s$) and is statistically significant. While all 2,543 recipes are included for enumeration at cut-off of two, only around 3% (80) and 0.6% (15) recipes are considered at cut-off of 15 and 20, respectively. While the recipes set controlled only for ingredient category did not explain the negative food pairing, controlling for frequency of use of ingredient reproduces the characteristic profile. (**B**) Strong negative food pairing emerged as an invariant feature of all sub-cuisines as measured in terms of average food pairing and its statistical significance.

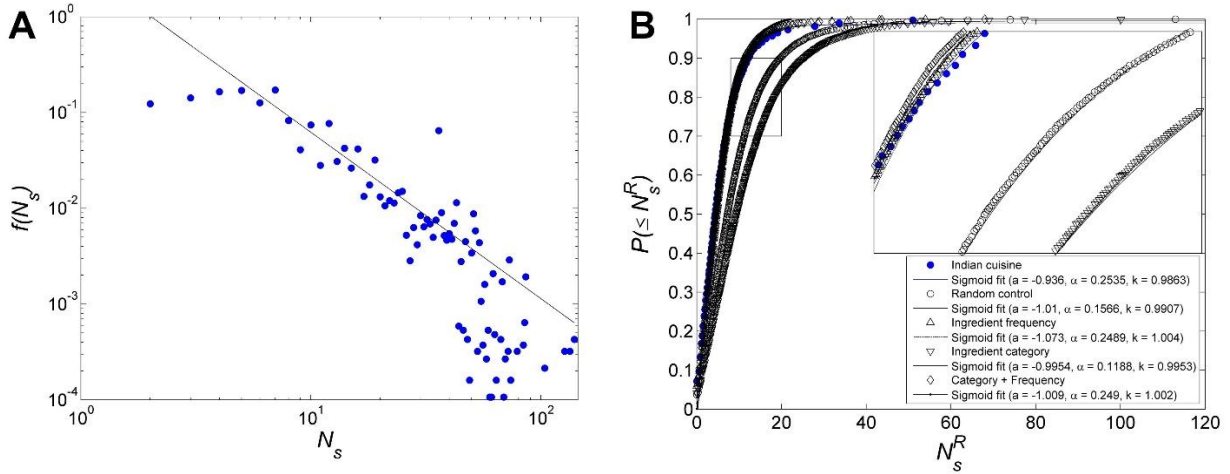

**Fig. 4. Negative food pairing at ingredient level and investigation of food pairing with recipe-level statistics.** (**A**) Fraction of ingredient pairs' frequency $f(N_s)$ with increasing number of shared flavor compounds ($N_s$). The Figure shows that more the flavor sharing between two ingredients, the less is their pairing in the cuisine. The frequency of ingredient co-occurrence falls as a power law (with an exponent of -1.74). (**B**) Cumulative distribution of 'average number of shared flavor compounds of recipes' in a cuisine. Cumulative distribution of $N_s^R$ values of recipes in Indian cuisine and its controls. Each of these data are best fitted with a Sigmoid equation ($P(\leq x) = a + \frac{(k-a)}{1+e^{-\alpha x}}$) indicating that $N_s^R$ frequency follows an exponential distribution. These results corroborate the observation that frequency of use of ingredients is a key contributor to the food pairing pattern.

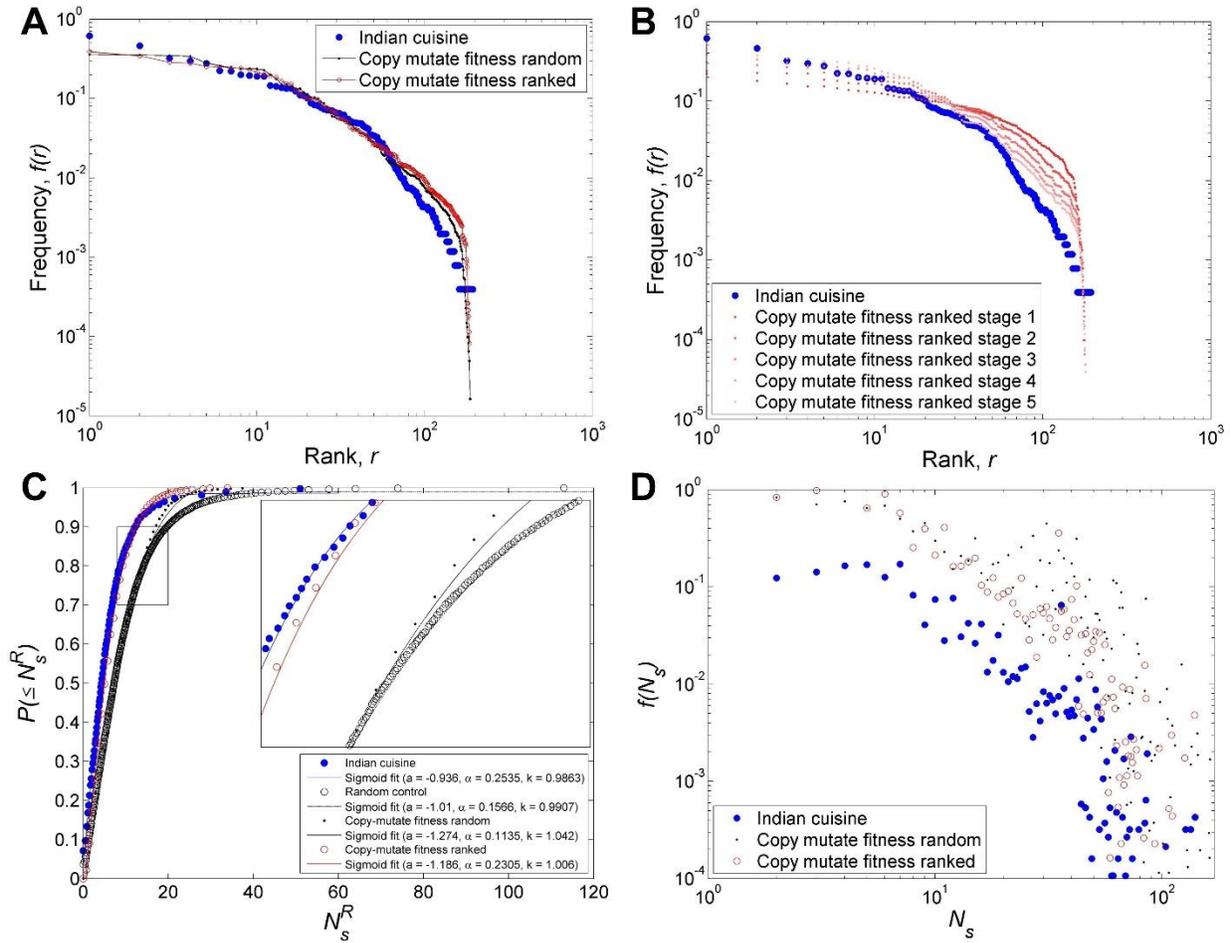

**Fig. 5. The modified copy-mutate model of Indian cuisine explains various aspects of negative food pairing.** (**A**) Frequency-rank distribution of ingredients generated by copy-mutate models of Indian cuisine. Both variants of the evolutionary model (one with random fitness value assignment and other with frequency scaled values of fitness) generated ingredient usage pattern closely matching that of Indian cuisine. The ingredient combinations of recipes generated by the latter model reflected the negative food pairing of the real-world cuisine. (**B**) The model suggests that the cuisine may have had similar pattern of rank distribution at different stages of its evolution. (**C**) The food pairing pattern is reproduced with the modified copy-mutate model. (**D**) The model also reflects the decline in frequency of ingredient pair occurrence with increasing overlap in their flavor profile.

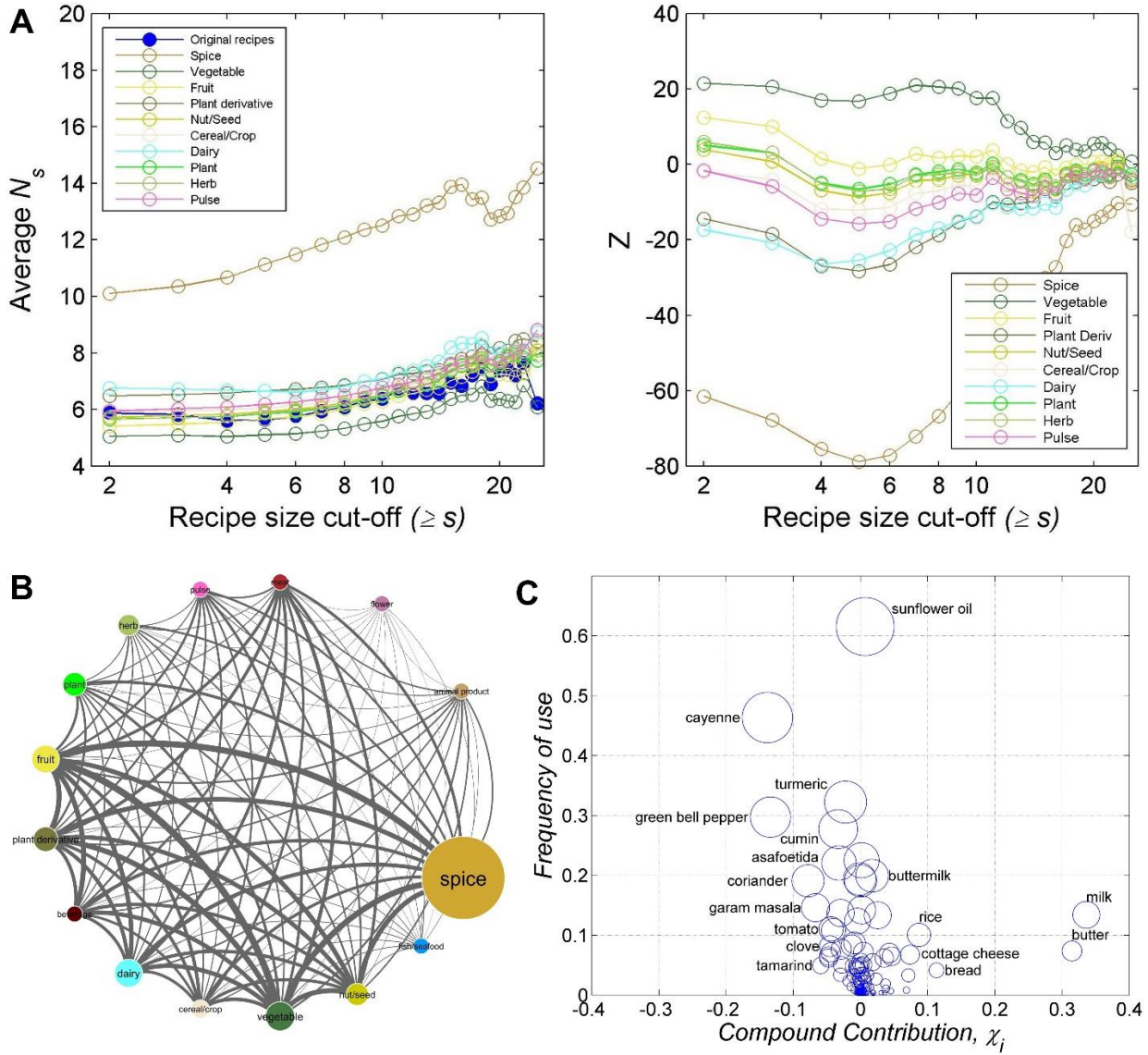

**Fig. 6. Spices are the critical contributors to the negative food pairing in Indian cuisine.** (**A**) Average food pairing of Indian cuisine when each ingredient of a given category is randomly replaced with another ingredient of the same category and its statistical significance. Such intra-category randomization reflects uniqueness of the ingredient in recipes knowing that ingredients tend to have similar flavor profiles within the category (Fig. S4). Spices are uniquely placed in the recipes, and when randomly replaced by another spice the flavor sharing pattern was drastically randomized. For a similar random intra-category replacement of ingredients of other categories, the flavor pattern showed no significant change. (**B**) Flavor sharing among ingredient categories. Size of circles denotes the extent of change that the category makes when its ingredients are randomly shuffled ($\Delta N_s^{cat}$), reflecting its importance in flavor sharing profile. (**C**) The relevance of individual ingredient enumerated in terms of extent of its contribution towards positive or negative food pairing ($\chi_i$) and frequency of use. Spices emerge as the most significant contributors to negative food pairing. Size of circles denote the frequency of use of ingredient.

# Supplementary Information

**Spices form the basis of food pairing in Indian cuisine**

Anupam Jain[a,†], Rakhi N K[b,†] and Ganesh Bagler[b,*]

## 1. Materials and methods

1.1 Data source selection

Our data on Indian cuisine was obtained from one of the leading cookery websites in India, tarladalal.com. Among the various online resources available for recipes from Indian cuisines, TarlaDalal (http://www.tarladalal.com) was found to be the best in terms of authentic recipes, cuisine annotations and coverage across major regional subtypes. The website has 3330 recipes from 8 Indian cuisines. Among others online sources, Sanjeev Kapoor (http://www.sanjeevkapoor.com) has 3399 recipes from 23 Indian cuisines. NDTV Cooks (http://cooks.ndtv.com) has 667 Indian recipes across 15 cuisines. Manjula's Kitchen (http://www.manjulaskitchen.com) is restricted to 730 Indian vegetarian recipes across 19 food categories. Recipes Indian (http://www.recipesindian.com) has 891 recipes from around 16 food categories. All Recipes (http://www.allrecipes.com) has only 449 recipes from 6 food categories. In comparison to these sources, Tarladalal.com is the best source of recipes for Indian cuisine. The statistics of sub cuisines, their recipes and ingredients is provided in Table S1.

1.2. Data collection and curation: Recipes

We started with the data containing 3330 recipes. This number was reduced after pruning to remove duplicates and other inconsistencies, to obtain 3037 recipes. Recipes containing any of the ingredients for which no flavor profile was available (349) were removed from the data, leaving us with 2688 recipes. For the purpose of flavor sharing analysis, we removed ingredients from 'snack' and 'additive' categories, as their flavor profiles could not be determined. Finally, all recipes having single ingredient were removed to obtain a final set of 2543 recipes.

The Indian cuisine from TarlaDalal comprise of 588 ingredients. For the purpose of mapping the ingredients to their flavor profiles (list of volatileflavor compounds present in that ingredient), these ingredient names were aliased to 339 source ingredients. These ingredients were categorised into seventeen ingredient categories based on the nature of ingredients. Out of these 339 aliased ingredients, we could obtain flavor profiles for 194 ingredients. By aliasing we mean, removing the redundant terms in ingredient names. For instance, *canned pineapple* was aliased as *pineapple*. Ingredient names in Hindi were translated to English. For instance *anardaana* was renamed as *pomegranate*. Further we aliased some of the ingredient names for the purpose of matching our ingredient names with their data. For instance *bay leaf* was named *bay laurel*. By crosschecking with dataset of Ahn *et al* (1), ingredients unique to Indian cuisine were identified. A full list of ingredient aliases, their categories and other details are provided in Dataset S1.

1.3. Compilation of the flavor profiles data

Our data of Indian cuisine as well as flavor compounds is not exhaustive. The original ingredient number had to be drastically reduced as flavor profiles of some of the ingredients could not be identified. Our main source of flavor compounds was obtained from the data made available by Ahn *et al*(1). Out of 194 ingredients with flavor profiles, 146 were obtained from dataset of Ahn *et al*(1). Ingredients for which the flavor data could not be obtained from either of these sources (31), flavor profile were compiled by extensive literature survey. All the flavor profiles were cross checked with those in $6^{th}$ edition (latest) of Fenaroli's Handbook of Flavor Compounds (2) so that their names are used consistently. Chemical Abstract Service numbers were also used as unique identifiers to bring consistency in nomenclature of flavor molecules. Detailed information of all the ingredients and their flavor profile are provided in Dataset S2.

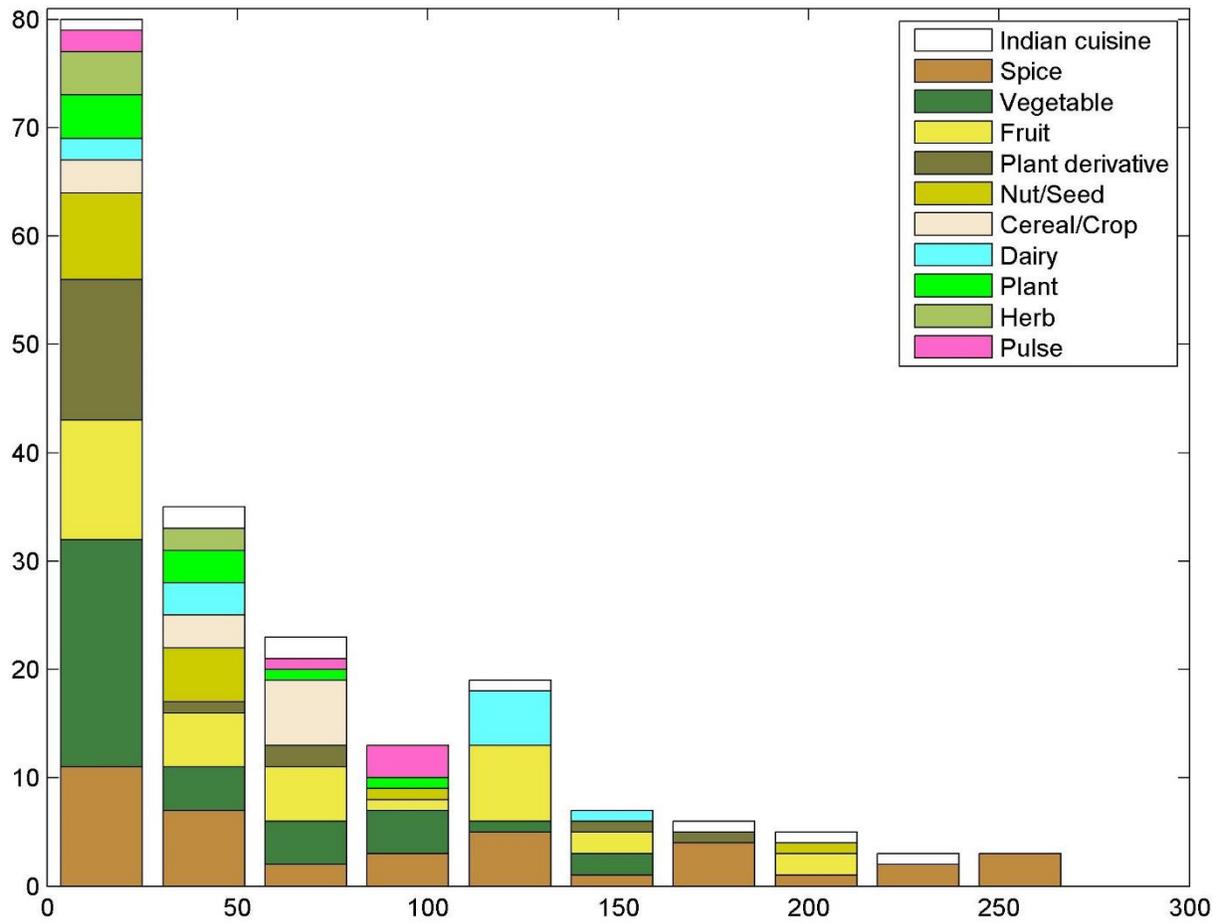

Fig. S1. **The flavor profile size distribution of ingredients from ten major categories**

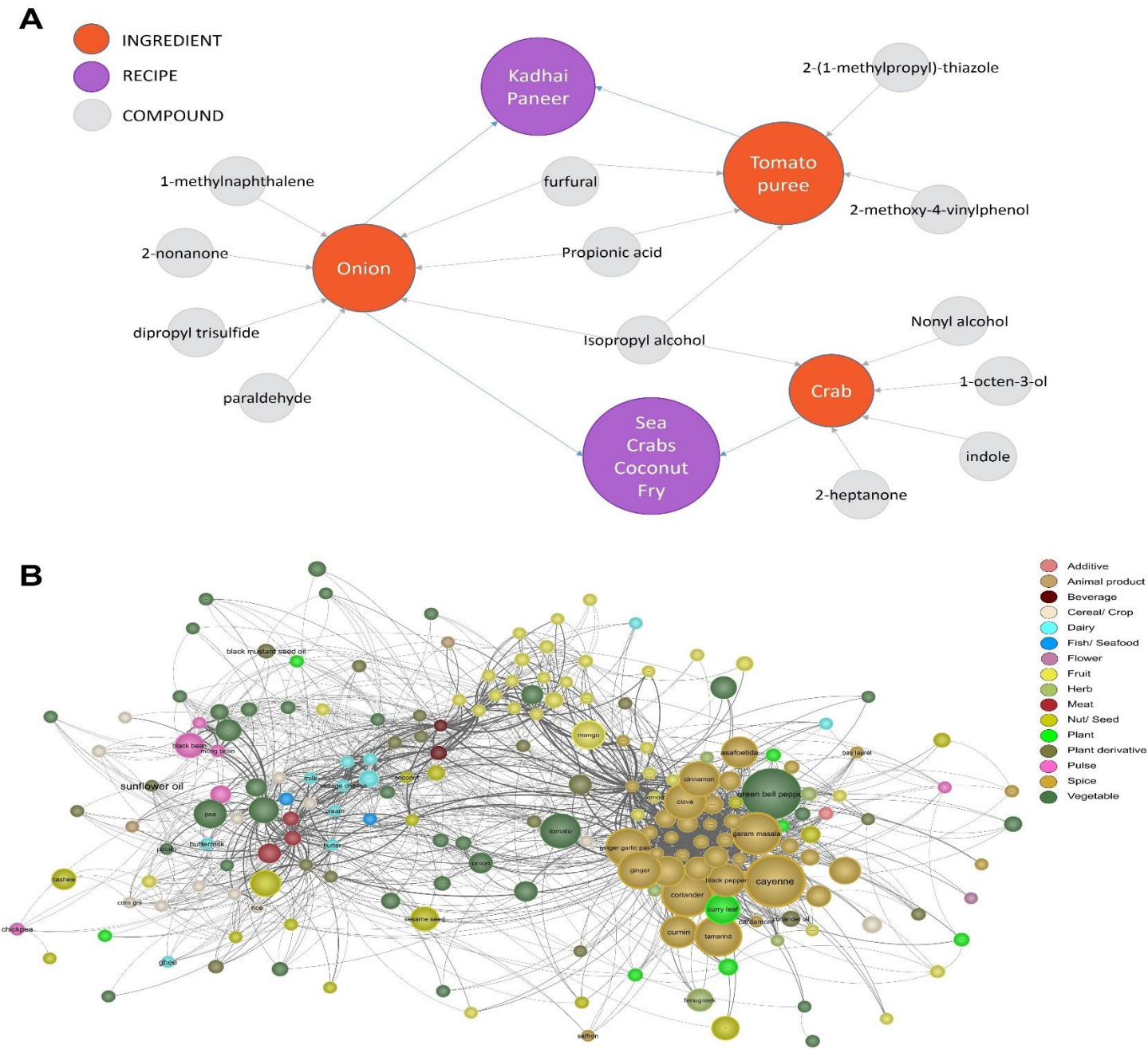

Fig. S2. **Construction of flavor graph.** (**A**) Illustration for construction of flavor graph of a cuisine starting from its ingredients set and their flavor profiles. (**B**) The backbone extracted (3) flavor graph of Indian cuisine. Ingredients are denoted by nodes and presence of shared flavor profile between any two ingredients is depicted as a link between them. The color of node reflects ingredient category and thickness of edges is proportional to extent of flavor profile sharing. Node size is scaled to the ingredient's contribution to negative food pairing of the cuisine

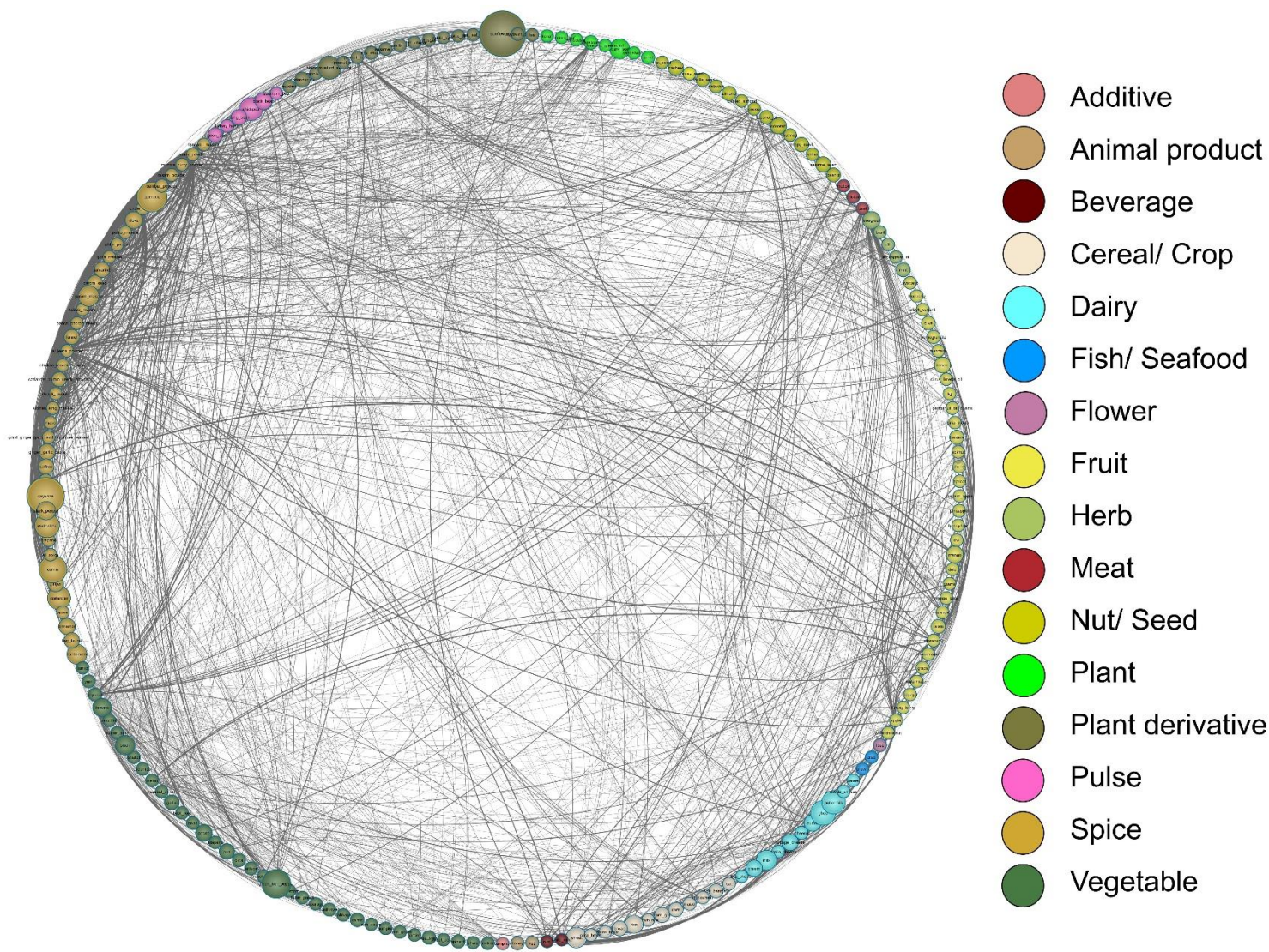

Fig. S3. **The backbone of flavor graph of Indian cuisine**. Each of the 194 ingredients is depicted as a node and shared flavor compounds are shown as edges. The size of the node is scaled to the frequency of use of the ingredient, whereas the thickness of the edge is scaled to number of shared flavor compounds.

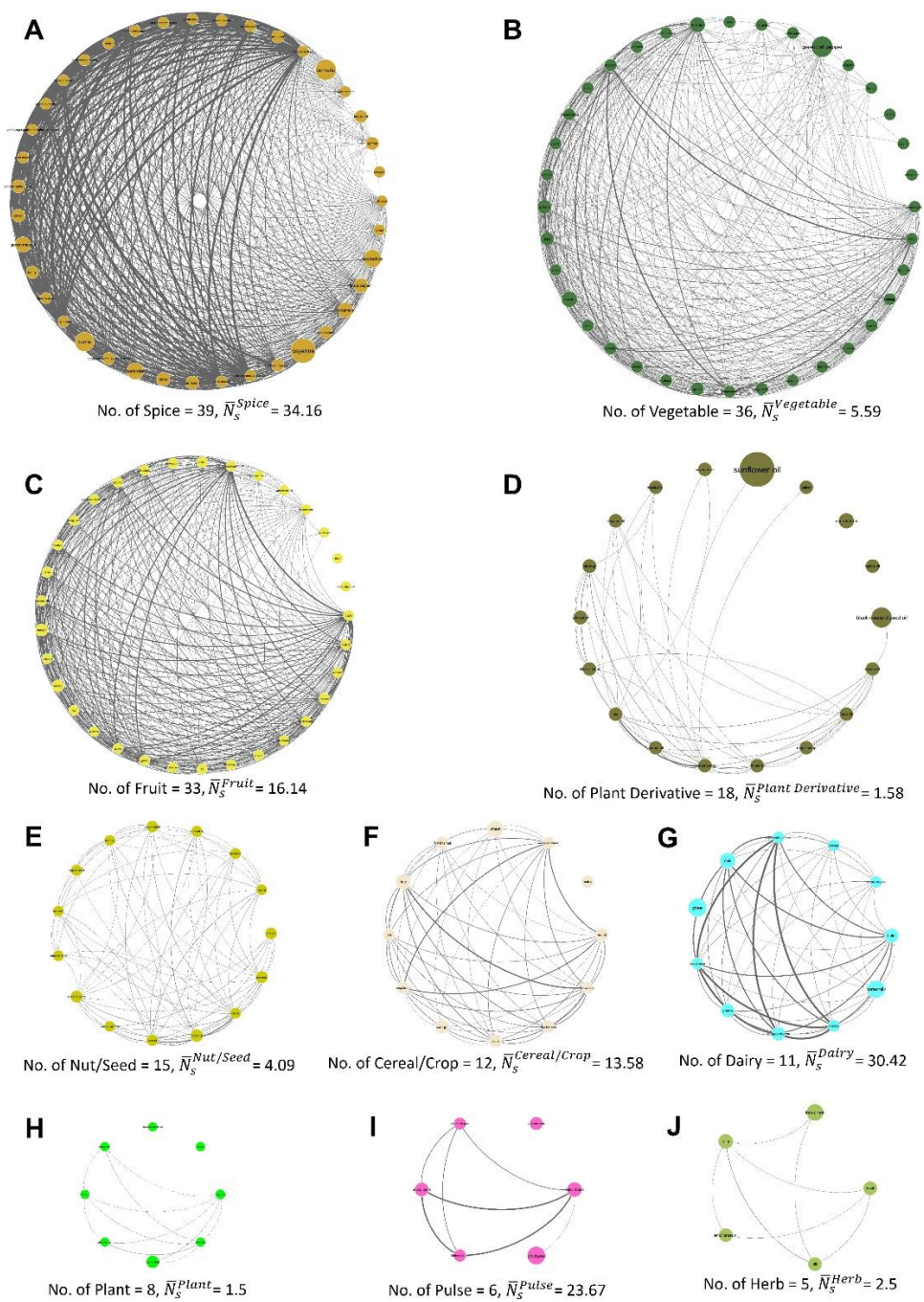

Fig. S4. **Flavor sharing within the ingredient category.** Intra-category flavor sharing pattern for 10 (of 15) major ingredient categories. The categories are color coded as per the legends in Figure S2B. (**A**) Spice, (**B**) Vegetables, (**C**) Fruit, (**D**) Plant derivative, (**E**) Nut/Seed, (**F**) Cereal/Crop, (**G**) Dairy, (**H**) Plant, (**I**) Pulse, (**J**) Herb

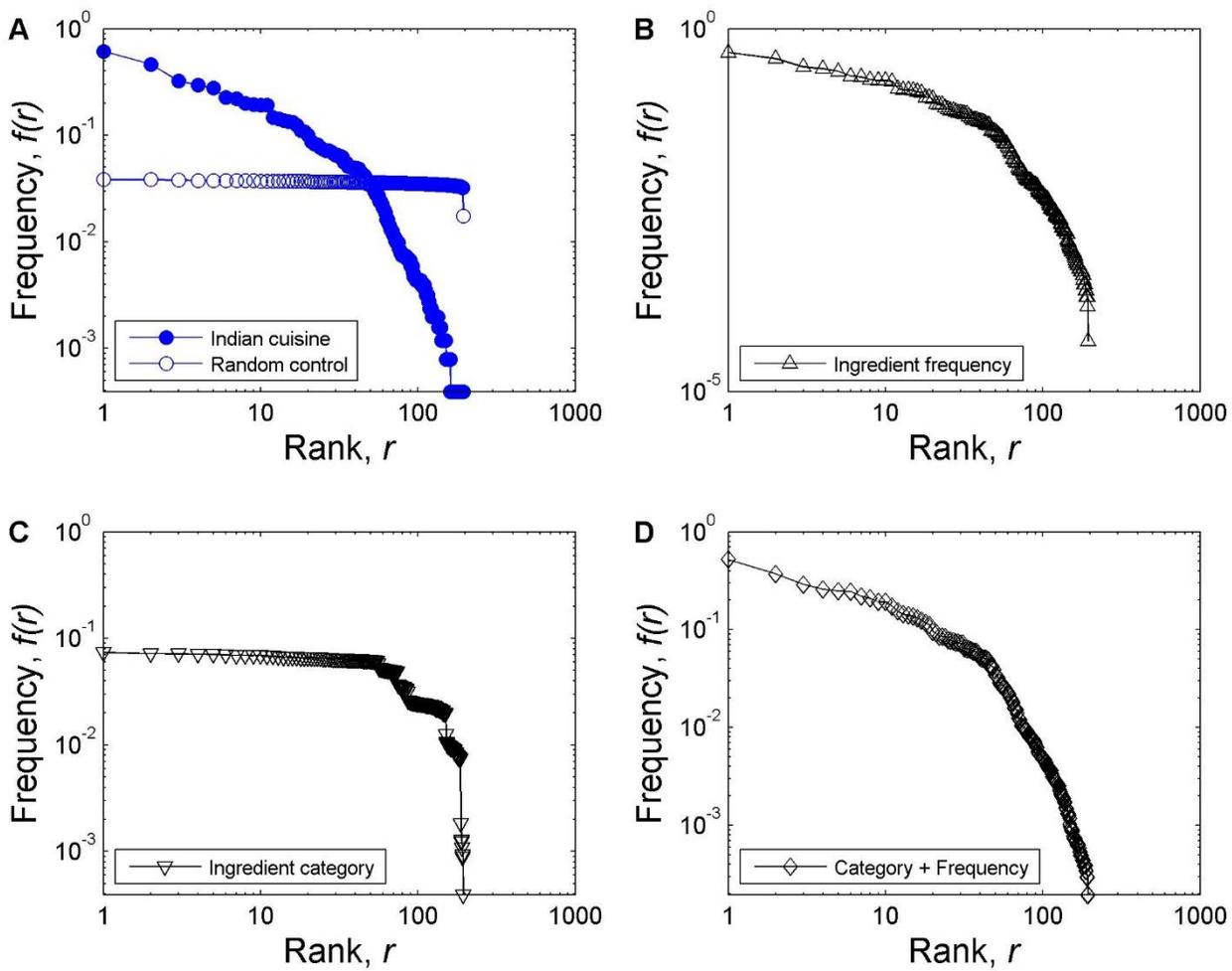

Fig. S5. **Ingredient rank profiles of Indian cuisine and controls.** (**A**) Random control when compared to Indian cuisine. (**B**) Control with ingredient frequency preserved. (**C**) Control that preserved only the ingredient category composition of a recipe. (**D**) Control in which, both, the frequency of use of ingredients as well as the category composition were preserved

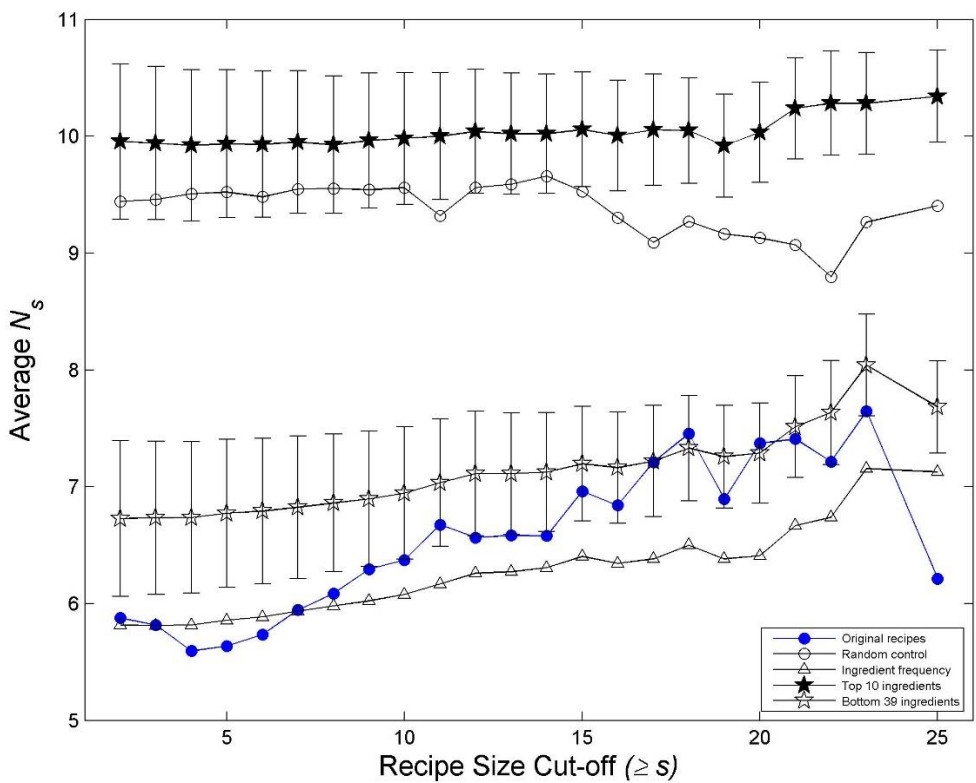

Fig. S6. **Role of most frequently used ingredients in the negative food pairing pattern of the Indian cuisine.** Frequency preserved random control with top 10 ranked ingredients swapped randomly with low-ranked ingredients exhibited food pairing pattern similar to a randomized cuisine. On the other hand, when poorly ranked ingredients (bottom 39 ingredients; equally ranked) were subjected to similar random swapping the food pairing was less affected. This highlighted that high-ranked ingredients are critical in specifying the characteristic profile of Indian cuisine. The error bars indicate the standard deviation over 10 experiments.

| Sub-Cuisine Name | No. of Recipes | No. of Ingredients (prior to aliasing) | No. of ingredients (post aliasing) |
|---|---|---|---|
| Bengali | 174 | 206 | 83 |
| Gujarati | 513 | 299 | 92 |
| Jain | 504 | 335 | 116 |
| Maharashtrian | 190 | 189 | 75 |
| Mughalai | 208 | 214 | 87 |
| Punjabi | 1085 | 365 | 119 |
| Rajasthani | 151 | 156 | 63 |
| South Indian | 581 | 265 | 92 |

Table S1. **Statistics of regional sub-cuisines that form the Indian cuisine**

| Sl. no | Ingredient category | No. of ingredients in the category |
|---|---|---|
| **1** | **Spice** | **39** |
| **2** | **Vegetables** | **36** |
| **3** | **Fruit** | **33** |
| **4** | **Plant derivative** | **18** |
| **5** | **Nut/seed** | **15** |
| **6** | **Cereal/crop** | **12** |
| **7** | **Dairy** | **11** |
| **8** | **Plant** | **8** |
| **9** | **Pulse** | **6** |
| **10** | **Herb** | **5** |
| 11 | Meat | 3 |
| 12 | Fish/seafood | 2 |
| 13 | Beverage | 2 |
| 14 | Animal product | 2 |
| 15 | Flower | 1 |

Table S2. **Number of ingredients in each category.** Ten major categories with five or more ingredients are highlighted.

| Sl.no | Ingredient name | Category | Frequency of occurrence |
|---|---|---|---|
| 1 | Ghee | Diary | 573 |
| 2 | Asafoetida | Spice | 561 |
| 3 | Garam masala | Spice | 372 |
| 4 | Curry leaf | Spice | 349 |
| 5 | Ginger garlic paste | Spice | 166 |
| 6 | Carom seed | Spice | 111 |
| 7 | Pigeon pea | Pulse | 90 |
| 8 | Coriander cumin seeds powder | Spice | 87 |
| 9 | Chat masala | Spice | 86 |
| 10 | Poppy seed | nut/seed | 83 |
| 11 | Rice basmati | cereal/crop | 69 |
| 12 | Nigella seed | nut/seed | 53 |
| 13 | Egg plant | Vegetable | 51 |
| 14 | Spinach | Vegetable | 41 |
| 15 | Pomegranate | Fruit | 38 |
| 16 | Sambar powder | Spice | 22 |
| 17 | Bitter gourd | Vegetable | 15 |
| 18 | Bottle gourd | Vegetable | 15 |
| 19 | Chole masala | Spice | 15 |
| 20 | Colocasia | Plant | 11 |
| 21 | *Pandanus fasicularis* | Fruit | 11 |
| 22 | Rasam powder | Spice | 11 |
| 23 | White pepper | Spice | 11 |

Table S3. **List of major ingredients not reported in other cuisines and are commonly used in Indian cuisine**

| Ingredients contributing to positive food pairing | χ value | Frequency of occurrence | Ingredients contributing to negative food pairing | χ value | Frequency of occurrence |
|---|---|---|---|---|---|
| Milk | 0.336059 | 341 | Cayenne | -0.13858 | 1179 |
| Butter | 0.314603 | 188 | Green bell pepper | -0.13416 | 756 |
| Bread | 0.113016 | 106 | Coriander | -0.07823 | 486 |
| Rice | 0.087081 | 256 | Garam masala | -0.06694 | 372 |
| Cottage cheese | 0.073573 | 172 | Tamarind | -0.05921 | 126 |
| Corn | 0.071018 | 84 | Ginger garlic paste | -0.04756 | 166 |
| Cheese | 0.068223 | 21 | Ginger | -0.04743 | 158 |
| Lemon | 0.046303 | 165 | Clove | -0.04557 | 208 |
| Grape | 0.044927 | 18 | Cinnamon | -0.04436 | 182 |
| Cream | 0.042721 | 179 | Tomato | -0.04381 | 281 |
| Honey | 0.037645 | 28 | Black pepper | -0.04037 | 275 |
| Olive | 0.037088 | 48 | Cumin | -0.03335 | 705 |
| Cocoa | 0.036144 | 10 | Asafoetida | -0.03201 | 561 |
| Coconut | 0.035244 | 158 | Coriander cumin seeds powder | -0.03032 | 87 |
| Strawberry | 0.030408 | 10 | Curry leaf | -0.02967 | 349 |

Table S4. **List of top 15 ingredients contributing to positive and negative food pairing**

| Sl.no | Ingredient name | Constituted spices | Freq of occurrence |
|---|---|---|---|
| 1 | Garam masala | Black pepper, mace, cinnamon, clove, cardamom, nutmeg | 372 |
| 2 | Ginger garlic paste | Ginger, garlic | 166 |
| 3 | Coriander cumin seeds powder | Coriander, cumin | 87 |
| 4 | Chaat masala | Asafoetida, mango, black salt, cayenne, garlic, ginger, roasted sesame seed, black mustard seed oil, turmeric, coriander, bay laurel, star anise, fennel | 86 |
| 5 | Sambar powder | Pigeon pea, coriander, chickpea, cumin, black pepper, cayenne, ginger, fenugreek, turmeric | 22 |
| 6 | Chole masala | Cayenne, garlic, ginger, roasted sesame seed, black mustard seed oil, turmeric, coriander bay laurel, star anise, fennel | 15 |
| 7 | Rasam powder | Cayenne, pigeon pea, cumin, Coriander, black pepper, curry leaf | 11 |
| 8 | Tandoori masala | Garlic, ginger, clove, nutmeg, mace, cumin, coriander, fenugreek, cinnamon, cardamom, black pepper | 8 |
| 9 | Curry powder | Cardamom, cayenne, cinnamon, clove, coriander, cumin, fennel fenugreek, mace, nutmeg, black pepper, poppy seed, roasted sesame seed, saffron, tamarind, turmeric | 5 |
| 10 | Kitchen king masala | Bay laurel, ginger, cinnamon, Clove, black pepper, coriander, Fennel, cayenne | 5 |
| 11 | Panch phoron seeds | Fenugreek, nigella seed, cumin, Black mustard seed oil, fennel | 4 |
| 12 | Chicken masala powder | Bay laurel, ginger, cinnamon clove, black pepper, coriander, fennel, cayenne | 2 |

| 13 | Goda masala | Cardamom, cinnamon, clove, Bay laurel, roasted sesame seed, Coriander, roasted coconut, Cassia, white pepper, Black pepper | 2 |
| --- | --- | --- | --- |
| 14 | Madras curry powder | Cardamom, cayenne, cinnamon, Clove, coriander, cumin, fennel, fenugreek, mace, nutmeg, black pepper, poppy seed, saffron, tamarind, turmeric | 1 |
| 15 | Jal jeera powder | Black salt, mango, cumin, citric acid, mint, black pepper ginger, asafoetida | 1 |
| 16 | Kebab masala | Bay laurel, ginger, cinnamon, clove, black pepper, coriander, fennel, cayenne | 1 |
| 17 | Grind ginger garlic and coriander leaves | Ginger, garlic, coriander | 1 |
| 18 | Pulao masala | Black pepper, white pepper, clove, cumin, cinnamon, cardamom, coriander | 1 |
| 19 | Dabeli masala | Cayenne, coriander, cinnamon, clove, cumin | 1 |

Table S5. **List of derived ingredients which are combinations of spices.**

**Additional Datasets S1-S3 (separate files)**

Dataset1 Recipes and Ingredients data of Indian cuisine
Dataset 2 Ingredients and their respective flavor compounds
Dataset 3 Ingredient contributions to negative and positive food pairing